\documentclass
[aps,prd,nofootinbib,
superscriptaddress,preprintnumbers,
twocolumn
]{revtex4-1}

\usepackage{hyperref}
\usepackage{color}
\usepackage[dvipsnames]{xcolor}%
\usepackage{graphicx,subfigure}
\usepackage{siunitx}
\usepackage{amsmath,amssymb,amsfonts}
\usepackage{bm}
\usepackage{hyperref}
\usepackage{lineno}
\usepackage{soul}
\usepackage{enumitem}
\usepackage{changepage}
\usepackage{float}
\usepackage[normalem]{ulem}

\graphicspath{{./plot/}}

\def \be {\begin{equation}}
\def \ee {\end{equation}}
\def \bes {\begin{subequations}}
\def \ees {\end{subequations}}

\newcommand{\beq}{\begin{eqnarray}}
\newcommand{\eeq}{\end{eqnarray}}

\newcommand{\Eq}[1]{Eq.~\eqref{#1}}
\newcommand{\Fig}[1]{Fig.~\ref{#1}}
\newcommand{\rref}[1]{Ref.~\cite{#1}}

\def\cM{{\mathcal M}}
\def\cL{{\mathcal L}}
\def\cR{{\mathcal R}}

\def\EB{{\rm EM}}
\def\p{{\bf p}}

\def\sel{{\sigma_{\rm el}}}

\newcommand\sect[1]{\noindent \textbf{#1.}---}

\begin{document}

\title{ 
The effect of weak magnetic photon emission from quark-gluon plasma
}

\author{Jing-An Sun}
\affiliation{Institute of Modern Physics, Fudan University, Handan Road 220, Yangpu District, Shanghai,
200433, China}

\author{Li Yan}
\affiliation{Institute of Modern Physics, Fudan University, Handan Road 220, Yangpu District, Shanghai,
200433, China}
\affiliation{Key Laboratory of Nuclear Physics and Ion-beam Application (MOE), Fudan University, Shanghai 200433, China}
\affiliation{Shanghai Research Center for Theoretical Nuclear Physics, NSFC and Fudan University, Shanghai 200438, China}

\date{\today}

\begin{abstract}

We propose a novel effect that accounts for the photon emission from a quark-gluon plasma in the presence of a weak external magnetic field. Although the weak magnetic photon emission from quark-gluon plasma only leads to a small correction to the photon production rate, the induced photon spectrum can be highly azimuthally anisotropic, as a consequence of the coupled effect of the magnetic field and the longitudinal dynamics in the background medium. With respect to a realistic medium evolution 
containing a tilted fireball configuration, the direct photon elliptic flow from experiments is reproduced. In comparison to the experimental data of direct photon elliptic flow, 
the strength of the magnetic field during the evolution of quark-gluon plasma 
can be extracted. For the top energy of RHIC collisions, 
$|eB|$ is found no larger than a few percent of the pion mass square.

\end{abstract}

\maketitle

\sect{Introduction}
The nature of high temperature quantum chromodynamics (QCD) is the major focus for the high-energy heavy-ion experiments carried out at Relativistic Heavy-Ion Collider (RHIC) and the Large Hadron Collider (LHC). In these facilities, quark-gluon plasma (QGP), a fluid with color degrees of freedom, has been created~\cite{Shuryak:2014zxa,Busza:2018rrf}. Dynamical properties of QGP have been well studied in terms of 
the observed spectra of various particles. Quite remarkably, at the top energies of RHIC and the LHC, a large number of hadron observables 
were found compatible with the theoretical modeling of QGP using viscous hydrodynamics, 
even at a high precision level~\cite{Gale:2013da,Shen:2020mgh}. 

Albeit its extreme success,  hydrodynamical modeling 
cannot describe photon productions from heavy-ion experiments~\cite{Gale:2012xq,Reygers:2022crp,Blau:2023bvi}. 
At the top RHIC energies, in the low $p_T$ region, experimentally measured direct photon yields  ({\it i.e.}, photon yields excluding those from hadron decays) exceed the current theoretical predictions~\cite{PHENIX:2014nkk}. 
More importantly, in experiments 
the spectrum of direct photons 
can be as azimuthally anisotropic as pions, with in particular a large elliptic flow $v_2^\gamma$~\cite{PHENIX:2015igl,ALICE:2018dti}. 
From the theoretical 
modeling, however, direct photons are expected more isotropic~\cite{Shen:2013cca,Chatterjee:2008tp,Gale:2021emg,Paquet:2015lta}.
This is a consequence that 
photon radiations from QGP 
are dominantly from the early stages~\cite{Shen:2013cca}, 
during which momentum anisotropy has not been fully developed. 
The discrepancy in both yields and elliptic flow 
is often referred to as the  ``direct photon puzzle'' (cf. \cite{Gale:2012xq,Reygers:2022crp}).

In theoretical models, to incorporate a significant emission anisotropy for the direct photons is challenging~\cite{Gale:2014dfa,Gale:2021emg}. 
The presence of an external magnetic field, on the other hand,  provides a possible solution.  In high-energy heavy-ion collisions, as a consequence of the relativistic motion of ions, magnetic fields are generated with extremely strong field strength~\cite{Skokov:2009qp,Bzdak:2011yy,Voronyuk:2011jd,Deng:2012pc}, with $|eB|/m_\pi^2$ reaches $O(10)$ at the top energies of RHIC and $ O(10^2)$ at the LHC, where $m_\pi$ is the pion mass. 
Although the influence of strong magnetic fields has already driven a number of physical predictions of great interest~\cite{Kharzeev:2015znc,Huang:2015oca,Hattori:2016emy}, because the pre-equilibrium stage of the QGP, in which the magnetic field decays most drastically, is hardly conducting, magnetic fields are expected weak as the system starts to evolve hydrodynamically. 
For instance, at around 0.4 fm/c and in the center of the fireball, the residual strength of magnetic field can drop to $|eB|/m_\pi^2\sim O(10^{-2})$ in a non-central AuAu collision at the top RHIC energy. 
Nonetheless, after the pre-equilibrium stage, the detailed evolution of the magnetic fields in QGP remains undetermined, owing to the lack of knowledge of the electrical properties of the QGP medium~\cite{McLerran:2013hla,Tuchin:2013apa,Gursoy:2014aka,Yan:2021zjc,Stewart:2021mjz, Huang:2022qdn,Zhang:2022lje}.



Regarding photon productions, strong magnetic field assumption has been considered previously~\cite{Basar:2012bp,Bzdak:2012fr,Muller:2013ila,Tuchin:2014pka,Zakharov:2016mmc,Wang:2020dsr}, which indeed gives rise to anisotropic emission. For instance, the synchrotron radiation induced by a strong magnetic field presents naturally an elliptic mode~\cite{Tuchin:2014pka}. Note that the strong magnetic field assumption would modify the theoretical description dramatically. Especially, when $|eB|\gg T^2$, magnetohydrodynamics 
should be taken into account, while when $\sqrt{|eB|}\gg gT$, effect of 
Landau level excitations 
cannot be neglected in quark scatterings~\cite{Miransky:2015ava,Huang:2022fgq}. With respect to the realistic QGP system, these conditions lead to a rough criterion: $|eB|/m_\pi^2\sim O(1)$.
Nonetheless, since the strong magnetic field in QGP persists only in the pre-equilibrium stage and the pre-equilbrium space-time volumn fills only several percents of the entire QGP evolution, its influence to the photon radiation is suppressed.

In this Letter, we focus on the hydrodynamic stage of a QGP evolution, during which only a weak external magnetic field, $|eB|/m_\pi^2\ll 1$, remains along with the medium. 
In this weak field scenario, the bulk part of hydrodynamical modeling is not affected, whereas photon productions in QGP receive a small correction due to the magnetic field. This small correction, which we refer to as the effect of weak magnetic photon emission, results in a large anisotropy in the direct photon spectrum.


\sect{Weak magnetic photon emission}
Photons radiated from a thermalized QGP can be produced by $2\to2$ scattering processes among quarks and gluons ($1+2\to 3 + \gamma$)~\cite{Kapusta:2006pm}. In a kinetic theory approach, the production rate is~\cite{Gale:2021emg,Paquet:2015lta}
\allowdisplaybreaks
\begin{align}
\label{eq:rate}
\cR^\gamma
&=\frac{1}{2(2\pi)^3}\sum_i\int \frac{d^3 \p_1}{2E_1(2\pi)^3}
\frac{d^3 \p_2}{2E_2(2\pi)^3}\frac{d^3 \p_3}{2E_3(2\pi)^3}\nonumber\\
&\qquad\times (2\pi)^4\delta^4(P_1+P_2-P_3-P)|\cM_i|^2\nonumber\\
&\qquad\times f_1(P_1)f_2(P_2)[1\pm f_3(P_3)]\nonumber \\
&\approx \frac{40\alpha \alpha_s}{9\pi^2}\cL f_q(P) 
I_c\,,
\end{align}
where the summation is over the Compton and the quark-antiquark annihilation channels with respect to the scattering amplitutes $|\cM_i|^2$, and $f_1$, $f_2$ and $f_3$ are distribution functions of quarks and gluons, correspondingly. The last expression in \Eq{eq:rate} gives the rate in the small angle approximation~\cite{Berges:2017eom}, with $\cL$ a Coulomb logarithm, and $I_c= \int d^3 \p/(2\pi)^3 [f_g+f_q]/p$ effectively characterizing the conversion between a quark-antiquark and a gluon in the thermalized QGP~\cite{Blaizot:2014jna}. 

In the previous studies based on hydrodynamics, dissipative effects in the medium have been taken into account to the photon production~\cite{Paquet:2015lta}. These effects are introduced via viscous corrections to the the quark and gluon distribution functions, $\bar f=n_{\rm eq} + \delta f$, where $n_{\rm eq}$ is the equilibrium distribution and the correction $\delta f$ is linear in the shear or bulk viscosities.
Analogously,  a weak external electromagnetic field induces additional correction to the quark distribution function, $f_q= \bar f_q + f_\EB = f_q + \delta f_q + f_{\EB}$. At the leading order of $|eB|/T^2$, from a straightforward derivation in kinetic theory (a simple derivation is given in the Supplemental Materials and more detailed discussions can be found in Refs.~\cite{Puglisi:2014sha,Sun:2023rhh}), one finds,
\beq
\label{eq:feb}
f_{\EB}  = \frac{c}{8\alpha_\EB} \frac{\sel n_{\rm eq}(1-n_{\rm eq})}{T^3 p\cdot u}e Q_f F^{\mu\nu} p_\mu u_\nu  \,,
\eeq
where 
$eQ_f$ indicates the corresponding electrical charge of a quark, $\sigma_{\rm el}$ is the electrical conductivity and $u_\nu$ is flow four-velocity. 
Although \Eq{eq:feb} applies more generally to weak electro- and magnetic fields, in this Letter we only focus on the magnetic field components, $B_i=\epsilon_{ijk} F^{jk}$.  
Note that in a QGP medium with temperature above the crossover temperature $T_c$, $f_{\rm EM}/n_{\rm eq}\ll 1$ is guaranteed with respect to the weak 
field condition $|eB|/m_\pi^2\ll1$. 
\Eq{eq:feb} is consistent to the kinetic theory definition of charge current, $j_{\EB}^i  = \sel E^i =  e\sum_f g_f Q_f\int \frac{d^3\p }{(2\pi)^3 p^0} p^i f_{\EB}$, from which, depending on the number of quark flavors considered, the constant $c$ can be determined. 

Accordingly, the photon production rate receives corrections due to the presence of a weak external electromagnetic field, $\cR^\gamma = \bar{ \cR}^\gamma + \cR^{\gamma}_\EB$, with $\cR^\gamma_\EB$ linear in the field strength. 
The background rate $\bar{\cR}^\gamma$, which is entirely determined by $\bar f$, has been applied previously to calculate photon productions in heavy-ion collisions.
After a space-time integral with respect to the medium evolution, it leads to the photon invariant spectrum,
\be
\label{eq:bkg}
E_p\frac{d^3 \bar N}{d^3 \p}=\int_V\bar{\cR}^\gamma(P,X) = \bar v_0 (1 +  2 \bar v_2 \cos 2\phi_p)\,,
\ee
where $X$ contains the space-time dependence in terms of the proper time $\tau=\sqrt{t^2-z^2}$, transverse coordinates $x$, $y$ and space-time rapidity $\eta_s={\rm arctanh}(z/t)$. In this work, we take the beam axis along $z$, and $x$-$z$ plane is the reaction plane. 
In \Eq{eq:bkg} the Fourier decomposition of the invariant spectrum defines the yields $\bar v_0$ and elliptic flow $\bar v_2$ of direct photons from the background, respectively. 
Similarly, one has for the corrections due to a weak magnetic field,
\be
E_p\frac{d^3 N_{\EB}}{d^3 \p}=\int_V
{\cR}_\EB^\gamma(P, X) = v^\EB_0 (1 +  2  v^\EB_2 \cos 2\phi_p)\,.
\ee
Here, $v_0^\EB$ and $v_2^\EB$ should be understood as the additional yields  and elliptic flow of photons entirely associated with the corrections from the weak magnetic field. The final predictions of direction photon emissions are thereby
\be
\label{eq:v0v2}
v_0^\gamma = \bar v_0 + v^\EB_0\,,\quad
v_2^\gamma = \frac{\bar v_2 \bar v_0 + v_2^{\EB} v_0^{\EB}}{\bar v_0 + v_0^{\EB}}\,.
\ee 


Let us now explain the effect of weak magnetic photon emission from QGP. 
In the weak magnetic field scenario, the magnetic field is too weak to modify the perturbative QCD scattering processes~\cite{Huang:2022fgq}, but suffices to drive the medium slightly out of equilibrium. The shift in the momentum distribution of incoming quarks brings in an extra source of photon production on top of the $2\to2$ scatterings in both channels, which scales with temperture $T$ as $T^4$.

Unlike the background contribution, where the photon elliptic flow is accumulated according to the space-time evolution of the azimuthal geometry of the medium,  
anisotropy in the weak magnetic photon emission 
is generated from two coupled effects: (1) A weak magnetic field which is orientated out of reaction plane. (2) Longitudinal dynamics of the background medium. Especially for the elliptic emission, one needs a rapidity-odd dipolar moment in the space-time geometry of the background medium. To show this, one first notices that the emitted photon spectrum is largely determined by the quark distribution function (cf., \Eq{eq:rate}), namely, $\cR^\gamma_\EB\propto v_0^{\EB}+v_2^{\EB}v_2^{\EB}\cos 2\phi_p \sim f_\EB$. 
With respect to a magnetic field out of reaction-plane, $\vec B = B_y \hat y$, in the rate one expects from $F^{\mu\nu}p_\mu u_\nu \sim B_y p_x u_z \propto \cos \phi_p$. 
Therefore, to realize an elliptic emission which scales as $\sim\cos 2\phi_p$,  an extra dipolar moment $\cos\phi_p$ 
in the background quark distribution is required. Fortunately, 
in heavy-ion experiments this dipolar moment has already been confirmd. In fact, in terms of the observed rapidity-odd directed flow $v_1^{\rm odd}$~\cite{STAR:2008jgm,ALICE:2013xri,Chatterjee:2017ahy} and rapidity-even dipolar flow $v_1^{\rm even}$~\cite{Teaney:2010vd,STAR:2018gji} of charged hadrons, there exist both odd and even dipolar moments in the evolving medium. 
Moreover, as a consequence of the external electromagnetic field and the condition of local equilibrium, the rapidity-odd dipolar flow among charged quarks splits and the split satisfies $\Delta v_1 \sim v_1^{\rm odd}$~\cite{Gursoy:2014aka}.
Because the direct photons are measured in a symmetric rapidity window~\cite{PHENIX:2015igl,ALICE:2018dti}, 
only  the rapidity-odd dipolar moment contributes.

In the Supplemental Material, the effect of the weak magnetic photon emission is verified in the case of Bjorken flow, where $v_2^\EB=0.5$. 
Intuitively, this large elliptic flow can be understood since magnetic field is highly anisotropic in nature.

\sect{Hydrodynamical modeling with weak magnetic photon emission}
In non-central heavy-ion collisions, the medium created by the colliding nuclei exhibits asymmetric distribution in the longitudinal direction, due to partly the structure of nucleus and partly the effect of longitudinal fluctuations. Effectively, the asymmetry in the medium can be captured in terms of a tilted fireball, based on which, hydrodynamical modeling reproduces the experimentally measured directed flow $v_1^{\rm odd}$ of charge hadrons~\cite{Chatterjee:2017ahy}.

Following \rref{Chatterjee:2017ahy}, we take the initial entropy density distribution 
\begin{align}
\label{eq:s0}
s(\tau_0, \vec x_\perp, \eta_s) \propto w(\eta_s)[\chi N_{\rm coll}
&+ (1-\chi)(N^+_{\rm part} w^+(\eta_s) \nonumber\\ 
&+ N^-_{\rm part} w^-(\eta_s))]\,,
\end{align}
where $N_{\rm coll}$, $N_{\rm part}^+$ and $N_{\rm part}^-$ are the densities of binary collisions and participants of the forward and backward going nuclei, respectively. As in the standard Glauber model, entropy production receives contributions from binary collisions and participants, relatively determined by the constant $\chi$. Longitudinal description in \Eq{eq:s0} is introduced via the functions $w(\eta_s)$ and $w^\pm(\eta_s)$. The symmetric longitudinal profile,
\be
w(\eta_s) = \exp\left(-\theta(|\eta_s|-\eta_M)\frac{(|\eta_s|-\eta_M)^2}{2\sigma_\eta^2}\right)
\ee
accounts for the longitudinal spectrum of charged hadrons, while
\be
w^+(\eta_s) = 
\begin{cases}
0\,,\quad &\eta_s<-\eta_T\\
\frac{\eta_T+\eta_s}{2\eta_T}\,,\quad &-\eta_T\le \eta_s \le \eta_T\\
1\,, &\eta_s>\eta_T
\end{cases}
\ee
and $w^-(\eta_s) =  w^+(-\eta_s)$ give rise to rapidity-odd component. For a given collision centrality, the spatial geometry of the distribution relies entirely then on these parameters, $\eta_T$, $\eta_M$ and $\sigma_\eta$, which we choose as in \rref{Chatterjee:2017ahy}. Note in particular, $\eta_T$ determines the extent to which the fireball is tilted.

\begin{figure}
\begin{center}
\includegraphics[width=0.4\textwidth] {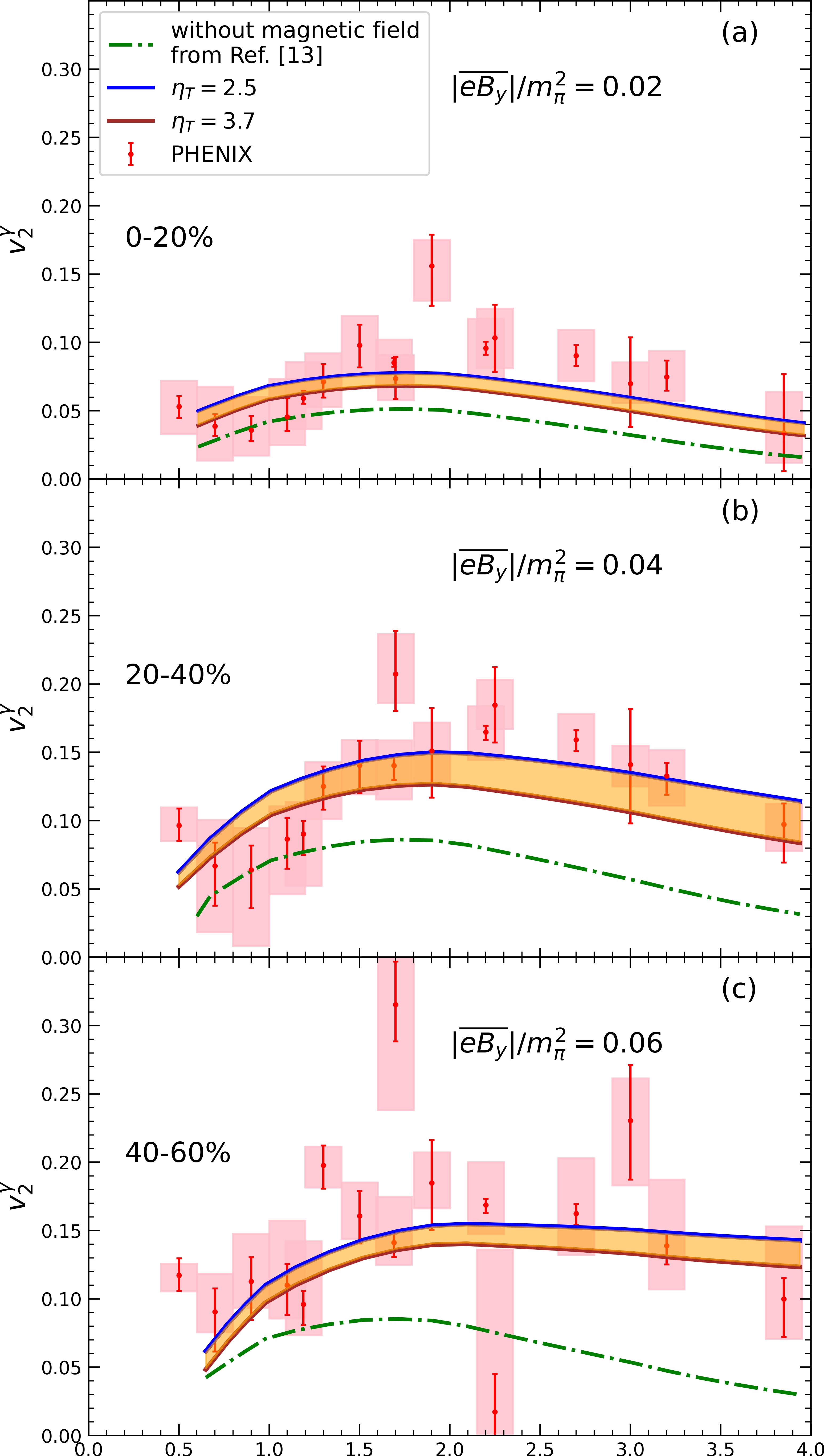}
\caption{
\label{fig:rhic_v2}
Direct photon elliptic flow at RHIC at different centralities. Green dash-dotted lines are from hydrodynamical modeling without the effect of external magnetic field~\cite{Gale:2021emg}. Final results with also weak magnetic photon emissions are shown as colored bands. Experimental data are from \rref{PHENIX:2015igl}.
}
\end{center}
\end{figure}

With respect to the initial condition \Eq{eq:s0}, we solve 3+1 dimensional viscous hydrodynamics using the state-of-the-art MUSIC program~\cite{Schenke:2010nt,Schenke:2010rr}, which has also been used for the calculation of the background direct photon spectrum for $\bar v_0$ and $\bar v_2$. To be consistent with the previous calculations in \rref{Gale:2021emg}, 
we consider the weak magnetic photon emissions from QGP between initial time $\tau_0=0.4$ fm/c and an effective crossover temperature $T_{c}=145$ MeV. 

After $\tau_0$, the QGP medium starts to evolve hydrodynamically, during which stage, with a finite electrical conductivity, 
the decay of magnetic field 
is expected much slower comparing to the vacuum field solution~\cite{Huang:2022qdn,Stewart:2021mjz}. Therefore, with respect to a QGP medium evolving hydrodynamically, we treat the magnetic field as a constant in our simulations, which allows us to estimate the averaged field strength throughout the QGP evolution. We consider the spatial distribution of the magnetic field according to the Lienard-Wiechert potential solution with respect to the moving nuclei up to $\tau_0$, namely, the vacuum field profile, following which we neglect the weak dependence on the transverse coordinates, while the dependence of the space-time rapidity is captured via a function, $\Gamma(\eta_s)$~\cite{Hattori:2016emy}. As a result, we have
\be
eB_y(\tau, \eta_s) = \overline{eB_y} \;\Gamma(\eta_s)\,,
\ee
where $\overline{e B_y}$ is a parameter effectively characterizing the time-averaged field strength of the system at the center of the fireball, between $\tau_0$ and the crossover temperature $T_c$.

\begin{figure}
\begin{center}
\includegraphics[width=0.4\textwidth] {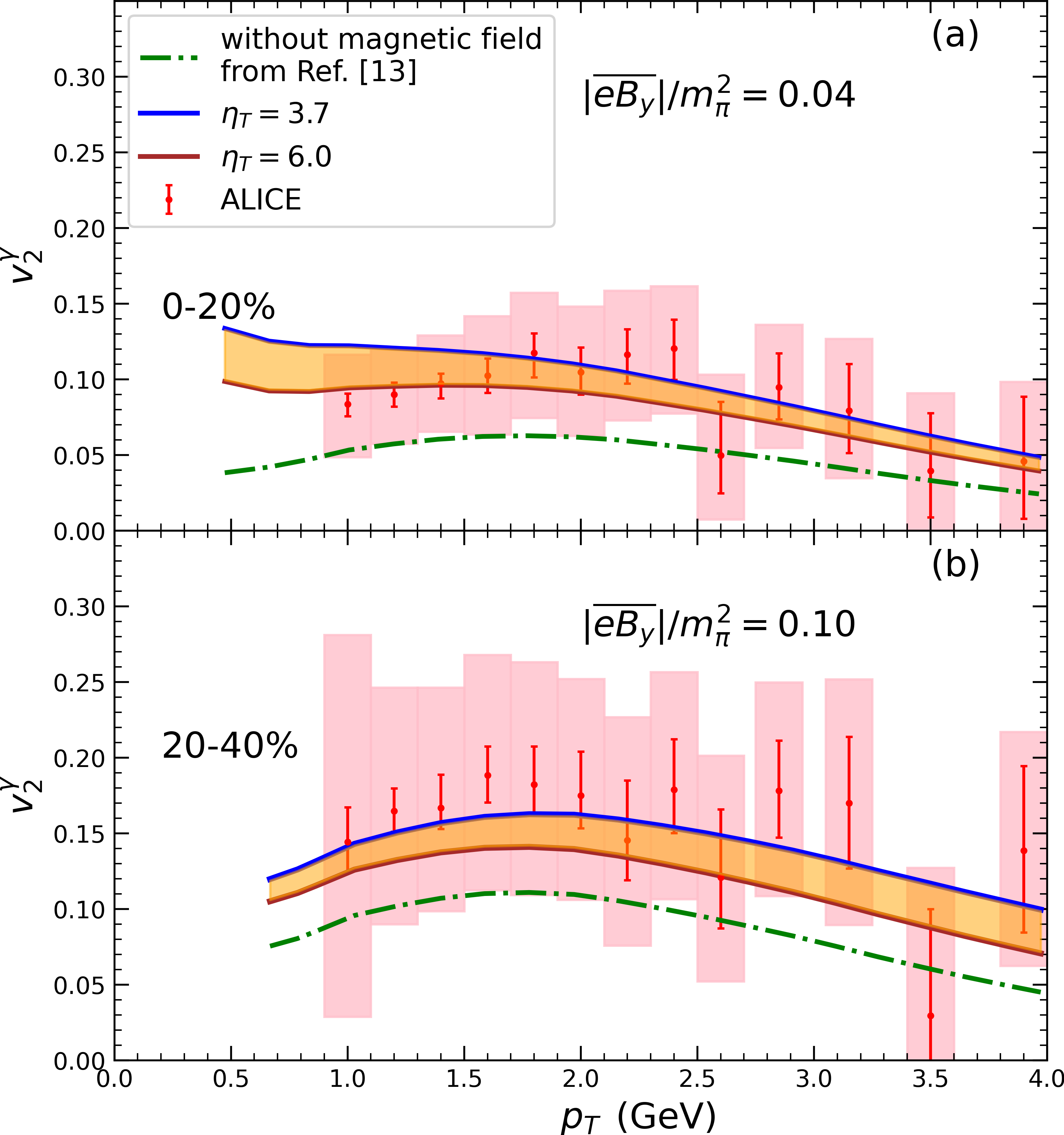}
\caption{
\label{fig:lhc_v2}
Direct photon elliptic flow at LHC at different centralities. Green dash-dotted lines are from hydrodynamical modeling without the effect of external magnetic field~\cite{Gale:2021emg}. Final results with also weak magnetic photon emissions are shown as colored bands. Experimental data are from \rref{ALICE:2018dti}.
}
\end{center}
\end{figure}

We will not calculate the 
yields $\bar v_0$ and elliptic flow $\bar v_2$ of direct photons from the background medium directly, instead we extract them from the most updated hydrodynamical modeling in \rref{Gale:2021emg}, 
where a variety of sources for photon emission have already been included. For instance, prompt photons produced from the initial hard scatterings are obtained via pQCD calculation at the NNLO order, photons from thermal radiations from QGP are calculcated with respect to the $2\to2$ scattering amplitudes determined via pQCD at the leading-log order~\cite{Arnold:2000dr}. 

To calculate $v_0^\EB$ and $v_2^\EB$, we consider $u$ and $d$ quarks that contribute to the photon emission. We take the small angle approximation for the photon production rate~\cite{Berges:2017eom,Churchill:2020uvk}, with respect to the magnetic field induced correction to the quark distribution function $f_\EB$ in \Eq{eq:feb}. To be consistent with the background calculations, we take the pQCD evaluation for the QGP electrical conductivity, $\sigma_{\rm el}/T\approx 5.98$~\cite{Arnold:2000dr,Huang:2022qdn}. With respect to the QGP evolution characterized by hydrodynamical modeling for a tilted fireball condition, we find $v^\EB_2\approx 0.6$, which is a bit larger than that from a simple Bjorken flow. Note that the value of $v^\EB_2$ does not depend on the magnitude of the magnetic field. 
Once $v_0^\EB$ and $v_2^\EB$ are given,
the yields and the elliptic flow of direct photons in heavy-ion collisions can be obtained according to \Eq{eq:v0v2}. 

\sect{Direct photon $v_2^\gamma$}
In \Fig{fig:rhic_v2}, the final results on the direct photon elliptic flow from RHIC AuAu collisions at $\sqrt{s_{NN}}=0.2$ TeV are shown for the corresponding three centrality classes. Comparing to the background contributions (green lines), with the weak magnetic photon emissions, the elliptic flow of direct photons gets enhanced.  
Quite remarkably, with the value of $\overline{eB_y}$ properly chosen, the resulted model prediction reproduces the experimental data. By doing so, we are allowed to extract the value of $\overline{eB_y}$. 
We find that as centrality grows, the extracted averaged field strength 
systematically increases, from $|\overline{eB_y}|=0.02 m_\pi^2$ at the 0-20\% centrality class, $|\overline{eB_y}|=0.04 m_\pi^2$ at the 20-40\% centrality class, to  $|\overline{eB_y}|=0.06 m_\pi^2$ at the 40-60\% centrality class. All these values satisfy the weak magnetic field condition, $|eB|/m_\pi^2\ll 1$. Weak magnetic photon emission leads to a small increase in the direct photon yields, which in the centrality class 20-40\%, is about 10\%. 

In \Fig{fig:lhc_v2}, the direct photon elliptic flow are shown similarly for the PbPb collisions at $\sqrt{s_{NN}}=2.76$ TeV. Comparing to the RHIC data, there exist large experimental uncertainties from the LHC measurements. Nevertheless, 
with the effect of weak magnetic photon emissions, 
the resulted elliptic flow is improved significantly. Following the same strategy, we extract the averaged field strength 
in the centrality classes 0-20\% and 20-40\%, leading to $|\overline{eB_y}|=0.04 m_\pi^2$ and $|\overline{eB_y}|=0.1m_\pi^2$, respectively.

We also investigate the effect of the background dipolar moment by varying the parameter $\eta_T$. As in \rref{Chatterjee:2017ahy}, we take $\eta_T$ approximately between 40\% of $y_{\rm beam}$ and $y_{\rm beam}-2.5$, so that the tilted fireball can capture the measured $v^{\rm odd}_1$ of charged hadrons. As expected, as shown as the colored bands in \Fig{fig:rhic_v2} and \Fig{fig:lhc_v2}, the effect of weak magnetic photon emission is stronger with respect to a larger dipolar moment.

\sect{Summary and discussion}We propose the weak magnetic photon emission as an extra source of photon productions from QGP,
due to the interplay between a weak magnetic field and the non-trivial longitudinal dynamics of the background QGP.  
In the cases of Bjorken flow (in Supplemental Material) and realistic 3+1 dimensional hydrodynamical simulations with respect to a tilted fireball, 
the effect of weak magnetic photon emission is verified.

As a novel source of photon production, the weak magnetic photon emission only leads to a small enhancement to the yields (about 10\%), while since the magnetic field is highly anisotropic in nature, the induced photons exhibit large elliptic flow ($v_2^\EB\sim 0.6$). Combined together, a finite increase in the elliptic flow of total direct photons is realized, which excellently explains experimental data.

In our simulations, all parameters, except 
$\overline{eB_y}$, are well determined according to the observed charged hadron spectrum. 
Therefore, by comparing the direct photon $v_2^\gamma$, we are allowed to estimate $\overline{eB_y}$, {\it i.e.} 
the time averaged field strength at the center of the fireball.
At the top RHIC energy and the LHC, the extracted field strength is only a few percent of the pion mass square.
In addition, 
we find a correct centrality dependence of the extracted 
field strength, as it increases towards peripheral collisions. 

Weak magnetic photon emission can be generalized to higher order flow harmonics of the direct photons. For instance, in a weak magnetic field, the longitudinally dependent elliptic moment in QGP would generate direct photon $v_3^\gamma$, while the longitudinal dynamics of a triangular moment can contribute to $v^\gamma_4$, etc. These non-trivial correlations between the longitudinal flow of charged hadrons and the spectrum of direct photons should be studied more systematically in future works, both theoretically and experimentally.
    
  \sect{Acknowledgements}We are grateful for very helpful discussions with Charles Gale and Xu-Guang Huang. This work is supported in part by the NSFC Grants through No.~11975079 and No.~12147101.
    
\bibliographystyle{apsrev}
\bibliography{ref}

\onecolumngrid
\newpage\hbox{}\thispagestyle{empty}

\section*{Supplemental Material}

\subsection{Derivation of \Eq{eq:feb}}

In analogy to the viscous corrections induced by shear and bulk viscous effects, dissipative correction to the quark distribution functions due to the external electromagnetic field can be derived. More detailed derivation can be found in Ref.~\cite{Sun:2023rhh}. For a system which is charge neutral locally, such as the QGP medium created in high-energy heavy-ion collisions, in the presence of an external electromagnetic field, the out-of-equilibrium effects can be solved via the Boltzmann-Vlasov equation. For instance, in the relaxation time approximation, 
\be
p^\mu \partial_\mu f +e Q F^{\mu\nu} p_\mu \frac{\partial f}{\partial p^\nu} = - \frac{p\cdot u}{\tau_R} \delta f\,.
\ee
The equation can be solved analytically by means of the Chapmann-Enskog method. With respect to the expansion of $\delta f$ in terms of $|eF^{\mu\nu}/T^2|$, one finds the leading order solution, 
\be
\delta f =  -\frac{\tau_R}{p\cdot u} e Q F^{\mu\nu}p_\mu \frac{\partial n_{\rm eq}}{\partial p^\nu}  = eQ F^{\mu\nu} p_\mu u_\nu \frac{\tau_R}{p\cdot u} \frac{n_{\rm eq}(1 - n_{\rm eq})}{T}\,.
\ee
Since the relaxation time $\tau_R$ is a parameter that characterizes interactions inside the system, it can be related to the transport coefficients.  Through the definition of the electrical charge current,
\be
j^\mu = \sigma_{\rm el} E^\mu = e \sum Q \int \frac{d^3 \p}{(2\pi)^3 E_p} p^\mu \delta f\,,
\ee
where the summation is over charged constituents and implicitly contains the degrees of freedom from color and spin, one finds $\sigma_{\rm el}\propto \tau_R T^2$. \Eq{eq:feb} can thus be obtained via a substitution of $\tau_R$ by $\sigma_{\rm el}$. The constant $c$
relies on the number of quark flavors. For a massless system with classical statistics, with 2-flavor quarks (anti-quarks), we have $c= 9\pi/10$, while for a system of 3-flavor quarks, $c=3\pi/4$.

\subsection{Bjorken flow}

For the purpose of illustration, we consider the background medium in terms of the Bjorken flow, namely, a flow pattern with longitudinal expansion which is  boost invariant and expansion in transverse directions is neglected. With the help of the Milne coordinates $(\tau,\eta_s)$, four-momentum and the flow four-velocity are
\begin{align}
p^\mu &= (p_T\cosh (Y-\eta_s), p_T\cos\phi_p, p_T \sin\phi_p, p_T \sinh(Y-\eta_s))\,,\nonumber\\
u^\mu &= (\cosh \eta_s,0,0,\sinh\eta_s)\,,
\end{align}
where $Y$ is the rapidity and $p_T$ the transverse momentum. Accordingly, in presence of an external magnetic field orientated along the $y$ direction, the correction in the quark distribution function owing to a weak magnetic field becomes,
\be
\label{eq:feb_simp}
f_{\EB}\propto 
 e Q_f B_y \frac{\sinh\eta_s }{ \cos(Y-\eta_s) }   n_{\rm eq}\cos\phi_p\,,
\ee
where for simplicity only factors of relevance are kept. 
There can be small anisotropic perturbations on top of the background medium, which are responsible for the anisotropic flow of the observed charged hadrons. In the similar manner as the tilted fireball, if only a rapidity-odd dipolar perturbation is included, one has
\be
n_{\rm eq} = A_0(\tau, \eta_s, p_T, Y) + A_1(\tau, \eta_s, p_T, Y)  \cos\phi_p \,,
\ee
where the explicit dependence in the functions $A_0$ and $A_1$ has been given. Note that, $A_0$ should be an even function of rapidity as it is related to the particle yields, while $A_1$ is an odd function in $Y$.
%
Substitute back to \Eq{eq:feb_simp}, one has
\begin{align}
\label{eq:bjk_feb}
f_{\EB} &\propto  Q B_y  \frac{\tau_R}{T}\frac{\sinh\eta_s }{ \cosh(y-\eta_s) }   (A_0 + A_1 \cos\phi_p )\cos\phi_p \cr
& = Q B_y  \frac{\tau_R}{T}\frac{\sinh\eta_s }{ \cosh(y-\eta_s) } \left[\frac{A_1}{2} + A_0 \cos\phi+
\frac{A_1}{2}\cos 2\phi \right]\,.
\end{align}
Since the spectrum of the weak magnetic photon emission is proportional to a space-time integral with respect to $f_\EB$,  \Eq{eq:bjk_feb} already implies that $v_2^\EB=0.5$. Since the direct photons in experiments are measured in a symmetric rapidity window, $Y\in[-Y_M,Y_M]$, in the brackets in \Eq{eq:bjk_feb} the terms that is symmetric in rapidity ($A_0\cos\phi_p$) should vanish, there is no extra $v_1^\EB$ contribution. This also explains why we only consider the rapidity-odd dipolar moment in this work.
 

\end{document}